# A mathematical study on the local fluid rotation axis

Charles Nottage[1], Yifei Yu[1], Chaoqun Liu[1*]

[1] Department of Mathematics, The University of Texas at Arlington, Arlington, Texas

*E-mail: cliu@uta.edu

**Abstract**: As widely recognized, vortex represents flow rotation. Vortex should have a local rotation axis as its direction and angular speed as its strength. Vorticity vector has been considered the rotation axis, and vorticity magnitude the rotational strength for a long time in classical fluid kinematics. However, this concept cannot stand in viscous flow. This study demonstrates by rigorous mathematical proof that the vorticity vector is not the fluid rotation axis, and vorticity is not the rotation strength. On the other hand, the Liutex vector is mathematically proved as the fluid rotation axis, and the Liutex magnitude is twice the fluid angular speed.

## 1. Introduction

Vortex is a widely observed phenomenon in nature and is of great significance to study the mechanism of turbulence and many other subjects. The physical properties of a vortex have been sought after by researchers for many decades. Vorticity was one of the first theories developed to identify the vortex. Helmholtz[1] suggested using vorticity tube/filament to represent vortex region. Researchers and textbooks have claimed that vorticity represents vortex[2]. Admittedly, vorticity can precisely derive rigid bodies' rotation, but the vortex prediction based on vorticity differs from experimental results. Robinson[3] discovered that the regions of strong vorticity and actual vortices were weakly correlated in turbulent boundary layers. Epps[4] also concluded that vorticity could not distinguish between a vortical region with rotational motions and a strong shear layer. The development of vortex identification methods spans decades, with many methods emerging to define and identify the vortical structure. One of the most

---

*Corresponding author: Chaoqun Liu, Professor, University of Texas at Arlington

E-mail: cliu@uta.edu

famous methods is the Q-criterion, proposed by Hunt et al.[5], using the residual of the squared vorticity tensor norm subtracted from the squared strain-rate tensor norm to represent vortex. Chong et al.[6] define vortex as the regions where the velocity gradient tensor $\nabla \vec{v}$ has one real and two complex conjugate eigenvalues, which is called $\Delta$ criterion. The $\lambda_{ci}$ method, proposed by Zhou et al.[7,8], is an extension of $\Delta$ criterion. These methods are scalar-valued, resulting in iso-surface, becoming the popular way to display the vortex structure. In this situation, the choice of threshold does not have a clear and universal criterion. Furthermore, the physical meaning of these methods is not clear; in other words, it is uncertain whether if the values obtained from these methods represent angular speed. These methods are unable to indicate the swirling axis as well. To solve these problems, Liu et al. proposed Liutex[9], a vector quantity, to capture vortex. The direction of Liutex represents the rotation axis, while its magnitude refers twice to the angular speed of the fluid rigid rotation. Afterward, Liu et al.[10] classified all vortex identification methods into three generations. Vorticity-based methods are considered the first generation. Velocity gradient eigenvalue-based methods, such as $Q^{[5]}$, $\lambda_{ci}$ [11], and $\lambda_2$ [12], are the second generation. Liutex-based methods are then the third generation. Liutex[9] is used as an indicator of vortex whose direction is the swirling axis and whose magnitude is twice angular speed of local rigid rotation. Although vortex theory has been in development for many years, there is yet a widely accepted mathematical definition of the vortex swirling axis. In Ref [19], Zhou et al. elaborated the hydrodynamic instabilities induced turbulent mixing in wide areas including inertial confinement fusion, supernovae and their transition criteria. Zhou detailed described Rayleigh–Taylor and Richtmyer–Meshkov instability and some related models in Ref. [20],[21]. Zhou's analysis is systematical, comprehensive, and sophisticated for flow instability and vortex generation, covering long history and state-of-the-art advances in turbulence research, which has clearly shown guidance for further and deeper scientific research. Liutex has the potential to help clarify the mechanism of these hydrodynamic instabilities as Liutex is a vector definition which more accurately represents flow rotation or vortex in comparison with Q or $\lambda_{ci}$ methods which are scalar and shear contaminated.

In this paper, some related concepts are reviewed in section 2; a mathematical definition of swirling axis is given from the physical essence of rotation axis, followed by the comparison of five candidate rotation axes in section 3. In section 4, a direct numerical simulation (DNS) result is taken as an example to illustrate the misconception of traditional theories. The mathematical misunderstanding of traditional vortex identification methods and the relations between some vortex identification methods are addressed.

## 2. Related concepts

First of all, some concepts on velocity tensor, velocity decomposition, and Liutex should be briefly reviewed.

**Definition 1:** Vorticity tensor

An arbitrary velocity gradient tensor can be decomposed using the traditional Cauchy-Stokes decomposition as follows:

$$\nabla \vec{v} = \begin{bmatrix} \frac{\partial u}{\partial x} & \frac{\partial u}{\partial y} & \frac{\partial u}{\partial z} \\ \frac{\partial v}{\partial x} & \frac{\partial v}{\partial y} & \frac{\partial v}{\partial z} \\ \frac{\partial w}{\partial x} & \frac{\partial w}{\partial y} & \frac{\partial w}{\partial z} \end{bmatrix} = A + B, \text{ where} \quad (2.1)$$

$$A = \begin{bmatrix} \frac{\partial u}{\partial x} & \frac{1}{2}\left(\frac{\partial u}{\partial y} + \frac{\partial v}{\partial x}\right) & \frac{1}{2}\left(\frac{\partial u}{\partial z} + \frac{\partial w}{\partial x}\right) \\ \frac{1}{2}\left(\frac{\partial v}{\partial x} + \frac{\partial u}{\partial y}\right) & \frac{\partial v}{\partial y} & \frac{1}{2}\left(\frac{\partial v}{\partial z} + \frac{\partial w}{\partial y}\right) \\ \frac{1}{2}\left(\frac{\partial w}{\partial x} + \frac{\partial u}{\partial z}\right) & \frac{1}{2}\left(\frac{\partial w}{\partial y} + \frac{\partial v}{\partial z}\right) & \frac{\partial w}{\partial z} \end{bmatrix}, \text{Symmetric part and} \quad (2.2)$$

$$B = \begin{bmatrix} 0 & \frac{1}{2}\left(\frac{\partial u}{\partial y} - \frac{\partial v}{\partial x}\right) & \frac{1}{2}\left(\frac{\partial u}{\partial z} - \frac{\partial w}{\partial x}\right) \\ \frac{1}{2}\left(\frac{\partial v}{\partial x} - \frac{\partial u}{\partial y}\right) & 0 & \frac{1}{2}\left(\frac{\partial v}{\partial z} - \frac{\partial w}{\partial y}\right) \\ \frac{1}{2}\left(\frac{\partial w}{\partial x} - \frac{\partial u}{\partial z}\right) & \frac{1}{2}\left(\frac{\partial w}{\partial y} - \frac{\partial v}{\partial z}\right) & 0 \end{bmatrix}, \text{Anti} - \text{Symmetric part} \quad (2.3)$$

The anti-symmetric part is known as the Vorticity tensor.

**Definition 2:** The Vorticity vector is defined as:

$$\vec{\omega} = \nabla \times \vec{v} = \left(\frac{\partial}{\partial x}, \frac{\partial}{\partial y}, \frac{\partial}{\partial z}\right)^T \times (u, v, w)^T = \left(\frac{\partial w}{\partial y} - \frac{\partial v}{\partial z}, \frac{\partial u}{\partial z} - \frac{\partial w}{\partial x}, \frac{\partial v}{\partial x} - \frac{\partial u}{\partial y}\right)^T \quad (2.4)$$

**Definition 3**: Liutex[9,13,14] $\vec{R}$ is a vector quantity that

$$\vec{R} = R\vec{r} \quad (2.5)$$

where $\vec{r}$ is the real eigenvector of velocity gradient tensor which satisfies $\vec{\omega} \cdot \vec{r} > 0$ and $R = \vec{\omega} \cdot \vec{r} - \sqrt{(\vec{\omega} \cdot \vec{r})^2 - 4\lambda_{ci}^2}$.

## 3. Mathematical definition of rotation axis and strength

In section 2, it has been shown that the direction of Liutex is the rotational axis, and the magnitude of Liutex is the vortex strength, which actually is twice the fluid angular speed.

Principal coordinate[15] is a special coordinate in which it is easy to decompose the velocity gradient tensor into rotation, shear, and stretching. From the process of constructing principal coordinate, it is easy to find the velocity gradient tensor has the form[15] of:

$$\nabla \vec{v} = \begin{bmatrix} \lambda_{cr} & -\frac{R}{2} & 0 \\ \frac{R}{2} + \varepsilon & \lambda_{cr} & 0 \\ \xi & \eta & \lambda_r \end{bmatrix} \quad (3.1)$$

where $R$ is the magnitude of Liutex, $\lambda_{cr}$ is the real part of the complex conjugate eigenvalues of the velocity gradient tensor and $\lambda_r$ is the real eigenvalue of the velocity gradient tensor. It is noted that principal coordinate is a local and instance concept, or, in other words, it can be different at different positions and at different instances of time. Note that we are only interested in the vortex region where $\nabla \vec{v}$ has one real and two conjugate complex eigenvalues. According to (2.5), $R = 0 \ if \ \lambda_{ci} = 0$ where $\nabla \vec{v}$ has three real eigenvalues.

**Definition 4**: Principal decomposition[15] is the decomposition in principal coordinate, i.e.

$$\nabla \vec{v} = \begin{bmatrix} \lambda_{cr} & -\frac{R}{2} & 0 \\ \frac{R}{2}+\varepsilon & \lambda_{cr} & 0 \\ \xi & \eta & \lambda_r \end{bmatrix} = \mathbf{R} + \mathbf{S} + \mathbf{C} \tag{3.2}$$

$$\mathbf{R} = \begin{bmatrix} 0 & -\frac{R}{2} & 0 \\ \frac{R}{2} & 0 & 0 \\ 0 & 0 & 0 \end{bmatrix}, \quad \mathbf{S} = \begin{bmatrix} 0 & 0 & 0 \\ \varepsilon & 0 & 0 \\ \xi & \eta & 0 \end{bmatrix}, \quad \mathbf{C} = \begin{bmatrix} \lambda_{cr} & 0 & 0 \\ 0 & \lambda_{cr} & 0 \\ 0 & 0 & \lambda_r \end{bmatrix}, \tag{3.3}$$

where **R** represents Rotation, **S** represents Shearing, and **C** represents Stretching.

Influenced by solid mechanics knowledge, people believed that anti-symmetric matrix **B** in (2.3) represents rotation. However, this belief does not match experimental results. For example, there is no rotation in Couette flow, but the anti-symmetric matrix **B** is not zero but large near the wall. Gao et al.[13] pointed out that the rotation matrix should be **R** which is the rotational part of anti-symmetric tensor **B**. **R** is Galilean invariant and independent of coordinate changes.

A rotational axis's fundamental physical property is that it can only be stretched (or compressed) along its length and cannot rotate itself or deform in other directions. Namely, the increment of velocity along the rotation axis must be along the rotational axis's direction.

**Definition 5:** At a moment, a local fluid rotation axis is defined as a vector that can only have stretching (compression) along its length.

This is a unique definition for rotation axis which any rotation axis must follow. It is basic math that the increment of $\vec{v}$ in the direction of $d\vec{r}$ is $d\vec{v} = \nabla\vec{v} \cdot d\vec{r}$ and **Definition 5** gives a rigorous mathematical definition of rotation axis that is $d\vec{v} = \nabla\vec{v} \cdot d\vec{r} = \alpha d\vec{r}$ along the rotation axis, which indicates $d\vec{r}$ is the real eigenvector of $\nabla\vec{v}$.

Any axis which does not meet **Definition 5** will not be a rotation axis.

We can figure out at least five candidate vectors that are considered as the possible rotation axis. They are the three real eigenvectors of **A**, vorticity, and Liutex. It will be found that only Liutex can meet **Definition 5**.

**Theorem 1:** Liutex is the local fluid rotation axis.

Proof. In the Liutex direction, which is the real eigenvector, $d\vec{v} = \nabla\vec{v} \cdot \vec{r} = \lambda_r \vec{r}$. According to **Definition 5**, Liutex is the Local rotation axis as $\vec{R} = R\vec{r}$.

**Theorem 2:** Vorticity is, in general, **not** local fluid rotation axis.

Proof. $d\vec{v} = \nabla\vec{v} \cdot \vec{\omega} = \mathbf{A} \cdot \vec{\omega} + \mathbf{B} \cdot \vec{\omega} = \mathbf{A} \cdot (a_1\vec{r_1} + a_2\vec{r_2} + a_3\vec{r_3}) + (\nabla \times \vec{v}) \times \vec{\omega}$

$$= a_1\lambda_1\vec{r_1} + a_2\lambda_2\vec{r_2} + a_3\lambda_3\vec{r_3} + 0 \neq \lambda(a_1\vec{r_1} + a_2\vec{r_2} + a_3\vec{r_3}) = \lambda\vec{\omega} \quad (3.4)$$

unless $\lambda_1 = \lambda_2 = \lambda_3 = \lambda$, where we assume $\vec{\omega} = a_1\vec{r_1} + a_2\vec{r_2} + a_3\vec{r_3}$

The velocity increment has stretched in three directions, which violates **Definition 5**, where $\vec{\omega} = (\nabla \times \vec{v})$, **A** and **B** are symmetric and anti-symmetric tensors of $\nabla\vec{v}$, $\lambda_1, \lambda_2, \lambda_3, \vec{r_1}, \vec{r_2}, \vec{r_3}$ are real eigenvalues and corresponding real eigenvectors of symmetric matrix **A** respectively. A rotation axis cannot be stretched in three directions but only in its own direction according to **Definition 5**.

**Theorem 3:** The eigenvectors of symmetric tensor **A** are not rotation axis.

Proof. Consider the velocity gradient tensor in principal coordinate and do the Cauchy-Stokes decomposition.

$$\nabla\vec{v} = \begin{bmatrix} \lambda_{cr} & -\frac{R}{2} & 0 \\ \frac{R}{2} + \varepsilon & \lambda_{cr} & 0 \\ \xi & \eta & \lambda_r \end{bmatrix} = \mathbf{A} + \mathbf{B} \quad (3.5)$$

$$\mathbf{A} = \begin{bmatrix} \lambda_{cr} & \frac{\varepsilon}{2} & \frac{\xi}{2} \\ \frac{\varepsilon}{2} & \lambda_{cr} & \frac{\eta}{2} \\ \frac{\xi}{2} & \frac{\eta}{2} & \lambda_r \end{bmatrix}, \quad \mathbf{B} = \begin{bmatrix} 0 & -\frac{R+\varepsilon}{2} & -\frac{\xi}{2} \\ \frac{R+\varepsilon}{2} & 0 & -\frac{\eta}{2} \\ \frac{\xi}{2} & \frac{\eta}{2} & 0 \end{bmatrix} \quad (3.6)$$

**A** is the symmetric part, and **B** is the vorticity tensor (anti-symmetric part).

The symmetrical tensor **A** has three real eigenvectors, where

$$\mathbf{A} = \begin{bmatrix} \lambda_{cr} & \frac{\varepsilon}{2} & \frac{\xi}{2} \\ \frac{\varepsilon}{2} & \lambda_{cr} & \frac{\eta}{2} \\ \frac{\xi}{2} & \frac{\eta}{2} & \lambda_r \end{bmatrix} \quad (3.7)$$

Let $\lambda_1$, $\lambda_2$, and $\lambda_3$ be the real eigenvalues of symmetric tensor **A**, and $\vec{d}_1$, $\vec{d}_2$ and $\vec{d}_3$ be the three corresponding eigenvectors of **A**, i.e., $\vec{d}_i = [x_i \quad y_i \quad z_i]^T$, for $i = 1,2,3$.

According to **Definition 5**, the definition of the velocity gradient tensor implies that,

$$d\vec{v} = \nabla\vec{v} \cdot d\vec{r} \quad (3.8)$$

Therefore,

$$\nabla\vec{v} \cdot d\vec{r} = \alpha d\vec{r} \quad (3.9)$$

If $\vec{d}_i = d\vec{r}$,

$$\nabla\vec{v} \cdot \vec{d}_i = \mathbf{A} \cdot \vec{d}_i + \mathbf{B} \cdot \vec{d}_i = \lambda_i \vec{d}_i + \mathbf{B} \cdot \vec{d}_i$$

$$= \lambda_i \begin{bmatrix} x_i \\ y_i \\ z_i \end{bmatrix} + \begin{bmatrix} 0 & -\frac{R+\varepsilon}{2} & -\frac{\xi}{2} \\ \frac{R+\varepsilon}{2} & 0 & -\frac{\eta}{2} \\ \frac{\xi}{2} & \frac{\eta}{2} & 0 \end{bmatrix} \cdot \begin{bmatrix} x_i \\ y_i \\ z_i \end{bmatrix}$$

$$= \lambda_i \begin{bmatrix} x_i \\ y_i \\ z_i \end{bmatrix} + \frac{1}{2} \begin{bmatrix} -(R+\varepsilon)y_i - \xi z_i \\ (R+\varepsilon)x_i - \eta z_i \\ \xi x_i + \eta y_i \end{bmatrix}$$

$$= \lambda_i \begin{bmatrix} x_i \\ y_i \\ z_i \end{bmatrix} + \left(-\begin{bmatrix} x_i \\ y_i \\ z_i \end{bmatrix} \times \begin{bmatrix} \eta \\ -\xi \\ R+\varepsilon \end{bmatrix}\right)$$

$$= \lambda_i \vec{d}_i + (-\vec{d}_i \times \vec{\omega}) \neq \alpha \vec{d}_i \quad (3.10)$$

This implies that $\vec{d}_i$ does not meet **Definition 5** and is thus not the rotation axis. This proves that the three eigenvectors of symmetrical tensor **A** are not rotation axi**s.**

Classical theory considers the vorticity vector as the rotation axis and vorticity magnitude as the vortex's strength, which is really a misunderstanding for over a century.

The other common misunderstanding is the consideration of $\lambda_{ci}$ as the fluid swirling/rotation strength.

**Theorem 4:** Liutex magnitude is the local rotation strength.

Proof. Liutex is the rigid rotation part extracted from the fluid motion and twice the local angular speed.

**Theorem 5:** $\lambda_{ci}$ is not the local rotation strength.

Proof. According to Zhou et al.[16]

$$\nabla \vec{v} = [\vec{v}_r \quad \vec{v}_{cr} \quad \vec{v}_{ci}] \begin{bmatrix} \lambda_r & 0 & 0 \\ 0 & \lambda_{cr} & \lambda_{ci} \\ 0 & -\lambda_{ci} & \lambda_{cr} \end{bmatrix} [\vec{v}_r \quad \vec{v}_{cr} \quad \vec{v}_{ci}]^{-1} = T\nabla\vec{V}T^{-1} \qquad (3.11)$$

The transformation is similar but not orthogonal since $T$ is not orthogonal. $\nabla\vec{V}$ has the same eigenvalues as original $\nabla\vec{v}$ has, but different rotation speed from $\nabla\vec{v}$ as vorticity of $\nabla\vec{v}$ is not equal to $\lambda_{ci} - (-\lambda_{ci}) = 2\lambda_{ci}$. This contradictory to the Galilean invariance of vorticity.

In the Liutex-based Principal Coordinate, $\nabla\vec{V}$ is precisely the same as the original $\nabla\vec{v}$ in the XYZ-frame since **Q** (transformation matrix) is the orthogonal matrix. Of course, other second-generation methods cannot be used as the fluid rotation strength either, like Q, Δ, or $\lambda_2$ which are all contaminated by shear and even have different dimensions from the fluid angular speed.

There is no cross-velocity increment perpendicular to the direction of the local rotational axis. For example, suppose the z-axis is the rotational axis in a reference frame. The velocity can only increase or decrease along the z-axis, which implies that $du=0$ and $dv=0$, but $dw$ is not necessarily 0.

**Theorem 2** has clearly shown vorticity vector, in general, is not rotation axis. The following theorem further confirms this conclusion.

**Theorem 6:** Vorticity, in general, is not the fluid rotation axis.

The vorticity vector

$$\nabla \times \vec{v} = \vec{\omega} = \begin{bmatrix} \eta \\ -\xi \\ R + \varepsilon \end{bmatrix} \qquad (3.12)$$

Since $\nabla\vec{v} \cdot d\vec{r} = d\vec{v} = \alpha d\vec{r}$, then $\nabla\vec{v} \cdot d\vec{r} = \alpha d\vec{r}$ if $d\vec{r}$ is the eigenvector of $\nabla\vec{v}$.

Let $d\vec{r} = \vec{\omega}$,

$$\nabla \vec{v} \cdot \vec{\omega} = \mathbf{A} \cdot \vec{\omega} + \mathbf{B} \cdot \vec{\omega} = \mathbf{A} \cdot \vec{\omega}$$

$$= \begin{bmatrix} \lambda_{cr} & \frac{\varepsilon}{2} & \frac{\xi}{2} \\ \frac{\varepsilon}{2} & \lambda_{cr} & \frac{\eta}{2} \\ \frac{\xi}{2} & \frac{\eta}{2} & \lambda_r \end{bmatrix} \cdot \begin{bmatrix} \eta \\ -\xi \\ R+\varepsilon \end{bmatrix} = \begin{bmatrix} \lambda_{cr} & 0 & 0 \\ 0 & \lambda_{cr} & 0 \\ 0 & 0 & \lambda_r \end{bmatrix} \cdot \begin{bmatrix} \eta \\ -\xi \\ R+\varepsilon \end{bmatrix} + \begin{bmatrix} 0 & \frac{\varepsilon}{2} & \frac{\xi}{2} \\ \frac{\varepsilon}{2} & 0 & \frac{\eta}{2} \\ \frac{\xi}{2} & \frac{\eta}{2} & 0 \end{bmatrix} \cdot \begin{bmatrix} \eta \\ -\xi \\ R+\varepsilon \end{bmatrix}$$

$$= \begin{bmatrix} \eta \lambda_{cr} \\ -\xi \lambda_{cr} \\ (R+\varepsilon)\lambda_r \end{bmatrix} + \frac{1}{2}\begin{bmatrix} R\xi \\ R\eta \\ 0 \end{bmatrix} = \lambda_{cr}\begin{bmatrix} \eta \\ -\xi \\ 0 \end{bmatrix} + \lambda_r\begin{bmatrix} 0 \\ 0 \\ (R+\varepsilon) \end{bmatrix} + \frac{R}{2}\begin{bmatrix} \xi \\ \eta \\ 0 \end{bmatrix} \tag{3.13}$$

Only in special cases, following equation can stand:

$$\lambda_{cr}\begin{bmatrix} \eta \\ -\xi \\ 0 \end{bmatrix} + \lambda_r\begin{bmatrix} 0 \\ 0 \\ (R+\varepsilon) \end{bmatrix} + \frac{R}{2}\begin{bmatrix} \xi \\ \eta \\ 0 \end{bmatrix} = \alpha\begin{bmatrix} \eta \\ -\xi \\ R+\varepsilon \end{bmatrix} \tag{3.14}$$

Following are two cases where vorticity could satisfy Definition 5:

1) No rotation: If $\lambda_r = \lambda_{cr}$ and $R = 0$ then,

$$\lambda_{cr}\begin{bmatrix} \eta \\ -\xi \\ 0 \end{bmatrix} + \lambda_{cr}\begin{bmatrix} 0 \\ 0 \\ (0+\varepsilon) \end{bmatrix} + \frac{0}{2}\begin{bmatrix} \xi \\ \eta \\ 0 \end{bmatrix} = \lambda_{cr}\begin{bmatrix} \eta \\ -\xi \\ \varepsilon \end{bmatrix} = \alpha\begin{bmatrix} \eta \\ -\xi \\ \varepsilon \end{bmatrix} \tag{3.15}$$

2) No shear related with $w$: If $\xi = \eta = 0$ then,

$$\lambda_{cr}\begin{bmatrix} 0 \\ 0 \\ 0 \end{bmatrix} + \lambda_r\begin{bmatrix} 0 \\ 0 \\ (R+\varepsilon) \end{bmatrix} + \frac{R}{2}\begin{bmatrix} 0 \\ 0 \\ 0 \end{bmatrix} = \lambda_r\begin{bmatrix} 0 \\ 0 \\ (R+\varepsilon) \end{bmatrix} = \alpha\begin{bmatrix} 0 \\ 0 \\ R+\varepsilon \end{bmatrix} \tag{3.166}$$

This implies that the vorticity vector is not the rotation axis in general. The vorticity vector becomes the rotation axis only when shear is zero, which is in general impossible for boundary layers unless the flow is inviscid. In Case 1), $\lambda_r = \lambda_{cr}$, $R = 0$, which has no rotation and no rotation axis, of course. Note that **Theorem 6** and **Theorem 2** are identical with different paths to prove.

Since the Liutex directional vector satisfies **Definition 5**, Liutex is the only candidate for the rotation axis. Note that since a normalized eigenvector is unique up to a ± sign, a second condition is imposed, $\vec{\omega} \cdot \vec{r} > 0$, where $\vec{\omega}$ is the vorticity vector[9].

4. **Contaminations of first and second generations of vortex identification**

Liutex is a physical quantity that exactly represents the rigid rotation or vortex, and all other methods are contaminated[17].

1) **Contamination of vorticity**

In a principal coordinate:

$$\vec{\omega} = (\eta, -\xi, R + \varepsilon)^T \qquad (4.1)$$

and its magnitude is

$$\|\vec{\omega}\| = \sqrt{\eta^2 + \xi^2 + (R + \varepsilon)^2} \qquad (4.2)$$

From equations (4.1) and (4.2), it can be concluded that a vorticity vector does not only represent rotation but also claims shear to be a part of the vortical structure.

2) **Contamination of the Q method**

In a principal coordinate:

$$\nabla \vec{V} = \begin{bmatrix} \lambda_{cr} & -\frac{1}{2}R & 0 \\ \frac{1}{2}R + \varepsilon & \lambda_{cr} & 0 \\ \xi & \eta & \lambda_r \end{bmatrix} = \begin{bmatrix} \lambda_{cr} & \frac{1}{2}\varepsilon & \frac{1}{2}\xi \\ \frac{1}{2}\varepsilon & \lambda_{cr} & \frac{1}{2}\eta \\ \frac{1}{2}\xi & \frac{1}{2}\eta & \lambda_r \end{bmatrix} + \begin{bmatrix} 0 & -\frac{1}{2}R - \frac{1}{2}\varepsilon & -\frac{1}{2}\xi \\ \frac{1}{2}R + \frac{1}{2}\varepsilon & 0 & -\frac{1}{2}\eta \\ \frac{1}{2}\xi & \frac{1}{2}\eta & 0 \end{bmatrix} = A_Q + B_Q \quad (4.3)$$

$$Q = \frac{1}{2}\left(\|B_Q\|_F^2 - \|A_Q\|_F^2\right) = \frac{1}{2}\left[2\left(\frac{R}{2} + \frac{\varepsilon}{2}\right)^2 + 2\left(\frac{\xi}{2}\right)^2 + 2\left(\frac{\eta}{2}\right)^2\right] - \frac{1}{2}\left[2\lambda_{cr}^2 + \lambda_r^2 + 2\left(\frac{\varepsilon}{2}\right)^2 + 2\left(\frac{\xi}{2}\right)^2 + 2\left(\frac{\eta}{2}\right)^2\right] = \left(\frac{R}{2}\right)^2 + \frac{1}{2}R \cdot \varepsilon - \lambda_{cr}^2 - \frac{1}{2}\lambda_r^2 \qquad (4.4)$$

Therefore, the value of Q is undoubtedly contaminated by shear and stretching. In addition, $Q$ contains an $R^2$ term, indicating dimensional inconsistency with a fluid rotation.

3) **Contamination of $\lambda_{ci}$ Criterion**

In a principal coordinate, we have

$$\frac{R}{2}\left(\frac{R}{2} + \varepsilon\right) = \lambda_{ci}^2 \qquad (4.5)$$

Thus,

$$\lambda_{ci} = \sqrt{\frac{R}{2}\left(\frac{R}{2} + \varepsilon\right)} \qquad (4.6)$$

The expression of $\lambda_{ci}$ has $\varepsilon$, which is a component of the shear part and thus is contaminated by shear. The result of a direct numerical simulation of boundary layer transition[18] is taken as an example to demonstrate our theories are correct. The point selected to analyze is shown in Fig. 1.

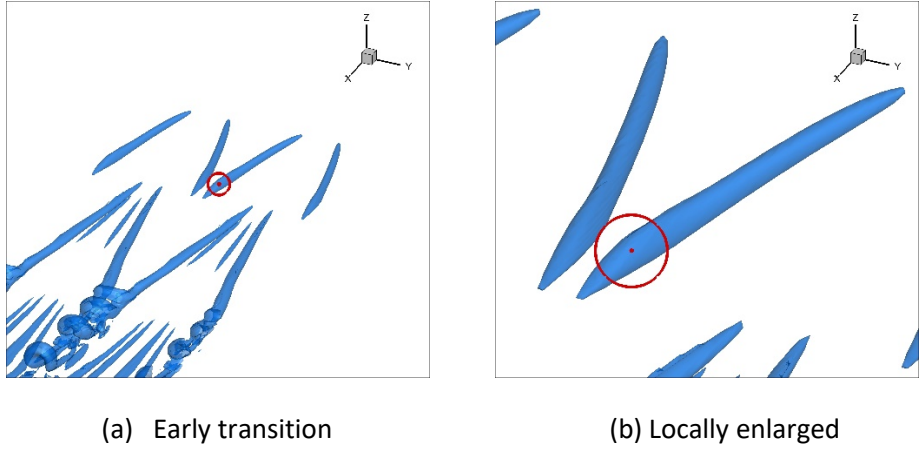

(a) Early transition  (b) Locally enlarged

Fig. 1 The selected point

The velocity gradient tensor at this point is

$$\nabla \vec{V} = \begin{bmatrix} 0.0533380 & 0.2818661 & 0.2621670 \\ -0.0139413 & 0.0003662 & 0.1193656 \\ -0.0055126 & -0.0798357 & -0.0548821 \end{bmatrix} \quad (4.7)$$

The corresponding Liutex and vorticity can be calculated as

$$\vec{R} = [-0.1292245 \quad 0.0197261 \quad -0.0100723]^T \quad (4.8)$$

$$\vec{\omega} = [-0.1992013 \quad 0.2676797 \quad -0.2958074]^T \quad (4.9)$$

The direction of the three eigenvectors of symmetric matrix D is:

$$\vec{d}_1 = [-0.6333458 \quad 0.4286110 \quad 0.6443335]^T \quad (4.107)$$

$$\vec{d}_2 = [0.1600420 \quad -0.7420691 \quad 0.6509377]^T \quad (4.118)$$

$$\vec{d}_3 = [0.7571391 \quad 0.5153891 \quad 0.4013906]^T \quad (4.129)$$

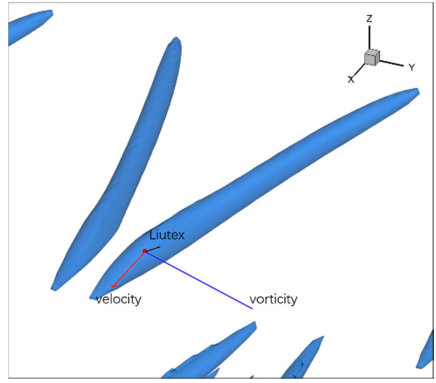

Fig. 2 Directions of Liutex, vorticity, and velocity

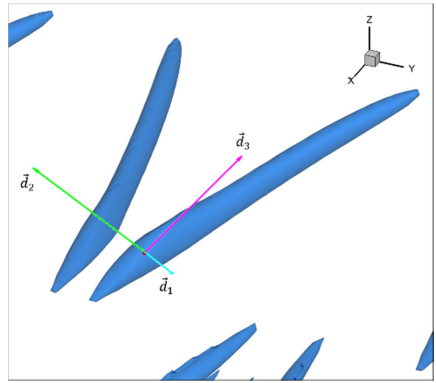

Fig.3 Directions of the three eigenvectors of the symmetric matrix

Fig. 2 shows the directions of Liutex, velocity, and vorticity. Fig.3 shows the directions of the three eigenvectors of the symmetric matrix **A**. It can be seen easily that the direction of Liutex is inside and generally parallel to the iso-surface shape while the other vectors are not. Besides the intuitive feeling from the graph, some mathematical analysis will be done. Recall that the rotation axis can only be stretched or compressed along its length, which is one of the physical essences of rotation. The increment of velocity along Liutex, vorticity, and eigenvectors can be described as follows.

$$d\vec{V}_R = \nabla \vec{V} \cdot \vec{R} = [-0.0039731 \quad 6.0649258 \quad -3.0968137]^T \quad (4.13)$$

$$d\vec{V}_\omega = \nabla \vec{V} \cdot \vec{\omega} = [0.0127261 \quad -0.0324341 \quad -0.0040377]^T \quad (4.14)$$

$$d\vec{V}_{d1} = \nabla \vec{V} \cdot \vec{d}_1 = [0.2559524 \quad 0.0858979 \quad -0.0660894]^T \quad (4.15)$$

$$d\vec{V}_{d2} = \nabla \vec{V} \cdot \vec{d}_2 = [-0.0299734 \quad 0.0751966 \quad 0.0226365]^T \quad (4.16)$$

$$d\vec{V}_{d3} = \nabla \vec{V} \cdot \vec{d}_3 = [0.2908864 \quad 0.0375455 \quad -0.0673495]^T \quad (4.17)$$

Whether the directions of two vectors are parallel or not can be tested by the result of cross-product. All vectors are normalized before taking the cross-product to avoid the influence of magnitudes.

$$(d\vec{V}_R)_{normed} \times \vec{R}_{normed} = [-3.5 \times 10^{-1} \quad 2.5 \times 10^{-17} \quad 1.1 \times 10^{-16}]^T \quad (4.18)$$

$$(d\vec{V}_\omega)_{normed} \times \vec{\omega}_{normed} = [0.68 \quad -0.19 \quad -0.63]^T \quad (4.19)$$

$$(d\vec{V}_{d1})_{normed} \times \vec{d}_{1\,normed} = [0.08 \quad -0.12 \quad 0.16]^T \quad (4.2010)$$

$$(d\vec{V}_{d2})_{normed} \times \vec{d}_{2\,normed} = [0.07 \quad 0.02 \quad 0.01]^T \quad (111)$$

$$(d\vec{V}_{d3})_{normed} \times \vec{d}_{3\,normed} = [0.05 \quad -0.17 \quad 0.12]^T \quad (12)$$

The definition of Liutex guarantees that its direction is parallel to the direction of velocity increment, and in our numerical analysis, the cross-product is almost zero. However, for the other four vectors, the cross-product result is far from zero, i.e., they do not satisfy **Definition 5** of the swirling axis.

The second condition for an axis to be the swirling axis is that it cannot rotate itself. In classical theory, vorticity tensor is misunderstood to represent the rotation part; however, the real rotation part should be described by Liutex. The rotation matrix $R$ is

$$R = \begin{bmatrix} 0 & -\frac{1}{2}R_z & \frac{1}{2}R_y \\ \frac{1}{2}R_z & 0 & -\frac{1}{2}R_x \\ -\frac{1}{2}R_y & \frac{1}{2}R_x & 0 \end{bmatrix} = \begin{bmatrix} 0 & 0.0050362 & 0.0098630 \\ -0.0050362 & 0 & 0.0646122 \\ -0.0098630 & -0.0646122 & 0 \end{bmatrix} \quad (4.23)$$

To avoid the influence of magnitudes, all vectors are normalized as well.

$$R \cdot \vec{R}_{normed} = [0 \quad 0 \quad 0]^T \quad (4.24)$$

$$R \cdot \vec{\omega}_{normed} = [-0.0035 \quad -0.0406 \quad -0.0344]^T \quad (4.25)$$

$$R \cdot \vec{d}_{1\,normed} = [0.0085 \quad 0.0448 \quad -0.0214]^T \quad (4.26)$$

$$R \cdot \vec{d}_{2\,normed} = [0.0027 \quad 0.0413 \quad 0.0464]^T \quad (13.27)$$

$$\boldsymbol{R} \cdot \vec{d}_{3\,normed} = [0.0066 \quad 0.0221 \quad -0.0408]^T \qquad (4.28)$$

Obviously, except Liutex, all other vectors rotate themselves.

The misunderstanding of some traditional vortex identification methods is explained below. People used to think vorticity $\vec{\omega}$ only has stretching along its direction and does not rotate itself because they thought $\vec{\omega}$ was only affected by the stretching matrix $\boldsymbol{A}$ since $\nabla\vec{v}\cdot\vec{\omega} = (\boldsymbol{A}+\boldsymbol{B})\cdot\vec{\omega} = \boldsymbol{A}\cdot\vec{\omega}$, and $\vec{\omega}$ does not rotate itself since $\boldsymbol{B}\cdot\vec{\omega} = 0$. This is a historical misunderstanding. Firstly, $\boldsymbol{A}$ represents stretching only when the direction is along one of its eigenvectors, and it does not work for an arbitrary direction, specifically the direction of vorticity. Eqs. 3.12-3.16 have shown that vorticity is the swirling direction only in some special cases. Vorticity is a good rotation indicator in solid mechanics because the shear is very small or zero, which is one of the special cases.

## 5. Conclusions

1) The vorticity tensor can be decomposed into a rigid rotational tensor and anti-symmetric shear tensor, which implies that vorticity is not strictly rotation. Instead, the vorticity vector contains shear and rotational factors.

2) The local rotational axis is defined as a vector along which the velocity increment can only have stretching (compression) along its length (**Definition 5**).

3) Liutex is the only candidate for local rotation axis as it satisfies **Definition 5**.

4) Vorticity vector is, in general, not the rotation axis, which directly opposes the traditional and classical concepts in fluid kinematics. Vorticity vector is rotation axis only when shear is zero, which cannot happen in boundary layers.

5) Vorticity magnitude is, in general, not the strength of fluid rotation unless shear is zero. Therefore, in general, vorticity vector is not vortex vector. They are two completely different and uncorrelated vectors.

6) Direction of vorticity used to be misunderstood as the rotational axis because people incorrectly consider matrix **A** means stretching, and anti-symmetric matrix **B** represents rotation.

7) The main problem of $\lambda_{ci}$ method is using similar transformations rather than orthogonal transformations. Therefore, it cannot guarantee the new space is isomorphic to the original one.

8) Vorticity and second generation of vortex identification methods are more or less contaminated by shear or stretching. Vorticity is the most contaminated, and that explains why Q and $\lambda_{ci}$ can detect vortex better than vorticity, but not as good as Liutex.

**Data availability**

The data that supports the findings of this study are available from the corresponding author upon reasonable request.

**Acknowledgments**

The authors thank the Department of Mathematics of the University of Texas at Arlington for the strong support. The authors are grateful to TACC (Texas Advanced Computation Center) for providing CPU hours to this research project. The computation is performed using Code DNSUTA, released by Dr. Chaoqun Liu at the University of Texas at Arlington in 2009.

**Reference**

[1] H. Helmholtz, "Über Integrale der hydrodynamischen Gleichungen, welche den Wirbelbewegungen entsprechen," 1435-5345 **1858** (55), 25 (1858).
[2] C. Truesdell, *The Kinematics of Vorticity(Indiana University Publications Science Seres Nr. 14.)* (Indiana University Press, Bloomington, 1954).
[3] S. K. Robinson, "Coherent Motions in the Turbulent Boundary Layer," Annu. Rev. Fluid Mech. **23** (1), 601 (1991).
[4] B. Epps, "Review of Vortex Identification Methods," in *AIAA SciTech Forum: 55th AIAA Aerospace Sciences Meeting* ([publisher not identified], [Place of publication not identified], 2017).
[5] J.C.R. Hunt, A.A. Wray, and P.Moin, "Eddies, stream, and convergence zones in turbulent flows," Center for turbulence research report CTR-S88, 193 (1988).
[6] M. S. Chong, A. E. Perry, and B. J. Cantwell, "A general classification of three-dimensional flow fields," Physics of Fluids A: Fluid Dynamics **2** (5), 765 (1990).
[7] J. ZHOU, R. J. ADRIAN, S. BALACHANDAR, and T. M. KENDALL, "Mechanisms for generating coherent packets of hairpin vortices in channel flow," J. Fluid Mech. **387**, 353 (1999).
[8] P. CHAKRABORTY, S. BALACHANDAR, and R. J. ADRIAN, "On the relationships between local vortex identification schemes," J. Fluid Mech. **535**, 189 (2005).

[9] C. Liu, Y. Gao, S. Tian, and X. Dong, "Rortex—A new vortex vector definition and vorticity tensor and vector decompositions," Physics of Fluids **30** (3), 35103 (2018).

[10] C. Liu, Y.-s. Gao, X.-r. Dong, Y.-q. Wang, J.-m. Liu, Y.-n. Zhang, X.-s. Cai, and N. Gui, "Third generation of vortex identification methods: Omega and Liutex/Rortex based systems," J Hydrodyn **31** (2), 205 (2019).

[11] J. ZHOU, R. J. ADRIAN, S. BALACHANDAR, and T. M. KENDALL, "Mechanisms for generating coherent packets of hairpin vortices in channel flow," J. Fluid Mech. **387**, 353 (1999).

[12] J. Jeong and F. Hussain, "On the identification of a vortex," J. Fluid Mech. **285** (-1), 69 (1995).

[13] Y. Gao, Y. Yu, J. Liu, and C. Liu, "Explicit expressions for Rortex tensor and velocity gradient tensor decomposition," Physics of Fluids **31** (8), 81704 (2019).

[14] Y.-q. Wang, Y.-s. Gao, J.-m. Liu, and C. Liu, "Explicit formula for the Liutex vector and physical meaning of vorticity based on the Liutex-Shear decomposition," J Hydrodyn **31** (3), 464 (2019).

[15] Y. Yu, P. Shrestha, C. Nottage, and C. Liu, "Principal coordinates and principal velocity gradient tensor decomposition," J Hydrodyn **32** (3), 441 (2020).

[16] Y. Gao and C. Liu, "Rortex and comparison with eigenvalue-based vortex identification criteria," Physics of Fluids **30** (8), 85107 (2018).

[17] V. Kolář and J. Šístek, "Stretching response of Rortex and other vortex-identification schemes," AIP Advances **9** (10), 105025 (2019).

[18] Y. Wang, Y. Yang, G. Yang, and C. Liu, "DNS Study on Vortex and Vorticity in Late Boundary Layer Transition," Commun. Comput. Phys. **22** (2), 441 (2017).

[19] Y. Zhou, T. T. Clark, D. S. Clark, S. Gail Glendinning, M. Aaron Skinner, C. M. Huntington, O. A. Hurricane, A. M. Dimits, and B. A. Remington (2019), "Turbulent mixing and transition criteria of flows induced by hydrodynamic instabilities," Physics of Plasmas **26** (8), 80901.

[20] Y. Zhou (2017), "Rayleigh–Taylor and Richtmyer–Meshkov instability induced flow, turbulence, and mixing. I," Physics Reports **720-722**, 1.

[21] Y. Zhou (2017), "Rayleigh–Taylor and Richtmyer–Meshkov instability induced flow, turbulence, and mixing. II," Physics Reports **723-725**, 1.
16 | 16